# Electrical control of near-field energy transfer between quantum dots and 2D semiconductors

*Dhiraj Prasai[†], Andrey R. Klots[#], AKM Newaz[#$], J. Scott Niezgoda[‡], Noah J. Orfield[‡], Carlos A. Escobar[§], Alex Wynn[#], Anatoly Efimov[∥], G. Kane Jennings[§], Sandra J. Rosenthal[†‡¶‡], Kirill I. Bolotin[#]\**

[†]Interdisciplinary Graduate Program in Materials Science, Vanderbilt University, Nashville TN, USA [#]Department of Physics and Astronomy, Vanderbilt University, Nashville TN, USA [$]Department of Physics and Astronomy, San Francisco State University, San Francisco CA, USA [‡]Department of Chemistry, Vanderbilt University, Nashville TN, USA [§]Department of Chemical and Biomolecular Engineering, Nashville TN, USA [∥]Center for Integrated Nanotechnologies, Los Alamos National Laboratory, Los Alamos NM, USA [¶]Vanderbilt Institute for Nanoscale Science and Engineering, Nashville TN, USA [‡]Materials Science and Technology Division, Oak Ridge National Laboratory, Oak Ridge TN, USA.

**ABSTRACT:** We investigate near-field energy transfer between chemically synthesized quantum dots (QDs) and two-dimensional semiconductors. We fabricate devices in which electrostatically gated semiconducting monolayer molybdenum disulfide ($MoS_2$) is placed atop a homogenous self-assembled layer of core-shell CdSSe QDs. We demonstrate efficient non-radiative Förster resonant energy transfer (FRET) from QDs into $MoS_2$ and prove that modest gate-induced variation in the excitonic absorption of $MoS_2$ lead to large (~500%) changes in the FRET rate. This, in turn, allows for up to ~75% electrical modulation of QD photoluminescence intensity. The hybrid QD/$MoS_2$ devices operate within a small voltage range, allow for continuous modification of the QD photoluminescence intensity, and can be used for selective tuning of QDs emitting in the visible-IR range.

**Introduction**

Nanoscale optical emitters – such as semiconductor quantum dots (QDs) or fluorophores - are strongly affected by their environment. An optical excitation in a nanoemitter can be transferred into the environment non-radiatively via processes such as charge transfer and Förster resonant energy transfer (FRET). Among these processes, FRET is a uniquely efficient long-range optical process.[1] Electrical control of FRET is desirable for potential applications of nanoemitters. To enable such control, materials with optical properties that respond to electric field are required. Recently discovered two-dimensional materials, such as graphene or MoS$_2$ are ideal for this purpose. Due to their atomic thickness, optical parameters of these materials can be controlled via electrostatic gating.[2-5] We therefore expect that by placing a nanoemitter onto a 2D material, it may be possible to electrically control the FRET pathway between the two systems.

Here, we explore FRET between chemically synthesized QDs and two-dimensional semiconductor (2DSC) monolayer molybdenum disulfide (MoS$_2$).[6,7] FRET in such a system is especially interesting due to the presence of tightly bound excitons in MoS$_2$ that are stable at room temperature.[8-11] Moreover, the oscillator strength of these excitons is strongly modified by the presence of the charge carriers in MoS$_2$.[2-4] We find strong quenching of photoluminescence (PL) for QDs near MoS$_2$, demonstrate that this quenching is due to FRET between QDs and excitons in MoS$_2$, and prove that other mechanisms such as charge transfer do not play a role in this system. Furthermore, we observe ~75% modulation of QD photoluminescence intensity with electrostatic gating of MoS$_2$. We find that this phenomenon is caused by ~500% electrical modulation of the QD/MoS$_2$ FRET rate. This, in turn, is due to changes in the near-field absorption of MoS$_2$ related to interaction of MoS$_2$ excitons with free charge carriers.

Very recently, related approaches have been demonstrated to achieve electrical control of the FRET rate for QDs and other nanoscale infrared emitters near another 2D material, graphene.[12,13] Our use of 2DSC offers several distinct advantages. The sizeable bandgaps of 2DSCs allow us to achieve electrical modulation of FRET from QDs emitting in the visible range. The strong electrical modulation of excitons in 2DSCs allows for the operation of devices with significantly reduced electrical fields, compared to graphene. Finally, we show selective modulation of QDs at desired wavelengths by choosing 2DSCs with corresponding excitonic features.[14]

**FRET between QDs and two-dimensional semiconductors**

To explore near-field energy transfer between QDs and 2DSCs, it is important to understand the conditions under which this type of transfer is expected. In general, FRET between two systems depends on their separation distance and the overlap integral between the absorption and emission spectra. The Fermi golden rule yields the following estimate for the FRET rate between a 0D and

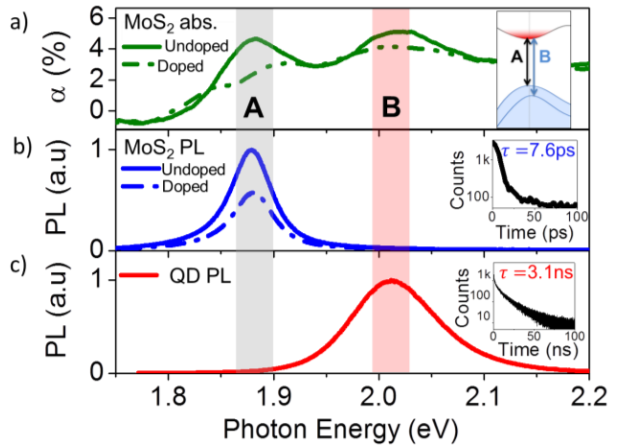

**Figure 1** (a) Absorption spectra of monolayer MoS$_2$ at two different doping levels. Inset: bandstructure of MoS$_2$ near its $K$-point. (b) PL spectra of monolayer MoS$_2$ at two different doping levels. Inset: time-resolved PL due to A-excitons in MoS$_2$. (c) PL spectrum of CdSSe QDs. Inset: time-resolved PL of excitons in QDs.

a 2D system (details in Supporting Information, S1):[1, 15, 16]

$$k_{FRET} \sim \frac{1}{d^4} \int_0^\infty \alpha(\lambda) f(\lambda) \lambda^4 d\lambda. \quad (1)$$

In this expression $f(\lambda)$ is the normalized emission of QDs, $\alpha(\lambda)$ is the absorption coefficient for a 2DSC as a function of wavelength $\lambda$, and $d$ is the distance between QDs and a 2DSC. The peculiar $d^{-4}$ dependence of $k_{FRET}$ is a characteristic of near-field coupling between excitations in 0D and 2D systems.[1, 16] Equation (1) indicates that in order to observe large $k_{FRET}$, the following conditions must be satisfied: *(i)* The optical absorption of the 2DSC must be sizable at the QD emission wavelength. *(ii)* A QD/2DSC separation $d$ must be small. *(iii)* The lifetime of an exciton in QDs, $\tau_{QD}$, must be longer than the inverse rate of energy transfer, $k_{FRET}^{-1}$. When this condition is fulfilled, an exciton in a QD lives long enough to transfer its energy into a 2DSC.

We can now select the appropriate materials to observe and explore FRET between QDs and 2DSCs. From the diverse group of 2DSCs (*e.g.*: $MoS_2$, $WSe_2$, $WS_2$), we chose monolayer $MoS_2$, a direct band gap semiconductor that is well studied, readily available, and optically active in the visible range.[6, 7] The absorption spectrum of $MoS_2$ (Fig. 1a) is dominated by two strong excitonic PL peaks at 1.88eV (A) and 2.05eV (B). These features are due to absorption of light by tightly bound band-edge A- and B-excitons[8-11] residing at the K-point of the Brillouin zone (Fig 1a, inset). The energy separation between the excitons is due to strong spin-orbit interaction[11] that splits the valence band of $MoS_2$. The photoluminescence spectrum of $MoS_2$ is dominated by A-excitons, the lowest excited state (Fig. 1b). With increased electron doping, both absorption (Fig. 1a, dashed line) and photoluminescence (Fig. 1b, dashed line) of $MoS_2$ are strongly reduced for energies corresponding to A- and B-peaks. This strong electro-optical effect is related to the interaction between excitons and free charge carriers in $MoS_2$. Doping-induced reduction of absorption is attributed to a combination of phase-space filling effect (blocking of low-momentum states that are needed for exciton formation) and screening of electron-hole interactions by free carriers.[17, 18] Additionally, doping allows the formation of charged

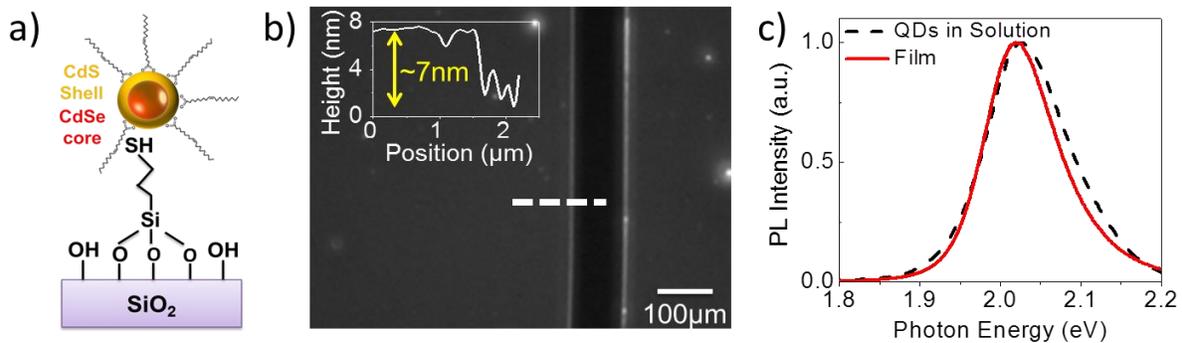

**Figure 2 (a)** CdSSe QDs with oleic acid ligands attached to functionalized $SiO_2$. **(b)** PL image of a QD film. A striation made on the film is evident as a dark strip. Inset: AFM height profile of the film obtained along the white dashed line in (b). **(c)** Normalized PL spectra of a QD film on $SiO_2$ and of the same QDs in solution.

excitons (trions),[3, 4] that become the new lowest-energy excitonic state and hence modify the PL spectrum.

We chose compositionally graded alloy core-shell CdSSe QDs[19] as the emission source. The QDs were synthesized to emit at ~2.02eV (Fig. 1c), very close to the B-peak in the absorption spectrum of $MoS_2$ (Fig. 1a). Additionally, CdSSe QDs are bright (quantum yield ~50%) and have lifetimes ~3ns (Fig. 1c, Inset). This is much longer than the ~8ps lifetime of excitons in $MoS_2$ (Fig. 1b, Inset; see "Methods" for measurement details). This ensures that FRET will be directed from QDs to $MoS_2$.[20, 21] Due to the spectral separation between the PL peaks of QDs and $MoS_2$, their spectra can be analyzed independently in hybrid structures.

Having spectrally satisfied FRET conditions in our hybrid structures, the next step is to physically bring QDs and a 2DSC in close proximity. We developed a flexible approach to address the biggest challenge in such devices – fabrication of uniform monolayer films of QDs. First, we used chemical self-assembly to deposit a uniform layer of QDs onto a $SiO_2$ substrate. The $SiO_2$ substrate functionalized with (3-Mercaptopropyl) trimethoxysilane was submerged into a solution of oleic acid-ligated CdSSe QDs (Fig 2a, see "Methods" for details).[22] The exposed thiol groups displace the oleic acid surface ligands and bind the QDs to the substrate.[23] The density of QDs was optimized to produce sub-monolayer films such that PL peaks due to QDs and $MoS_2$ could be distinguished. We used PL spectroscopy and atomic force microscopy (AFM) to assess the uniformity of QD films. With AFM we determined that the thickness of the QD film is ~7nm (Fig. 2b, Inset). This thickness is consistent with a sub-monolayer film of QDs that are ~5nm in diameter and have 1-2nm long oleic acid ligands.[24] Photoluminescence imaging indicates that as-fabricated QD films remain bright and are very uniform (Fig. 2b). Moreover, the position and the width of the PL peak for the QD film (Fig 2c, red line) do

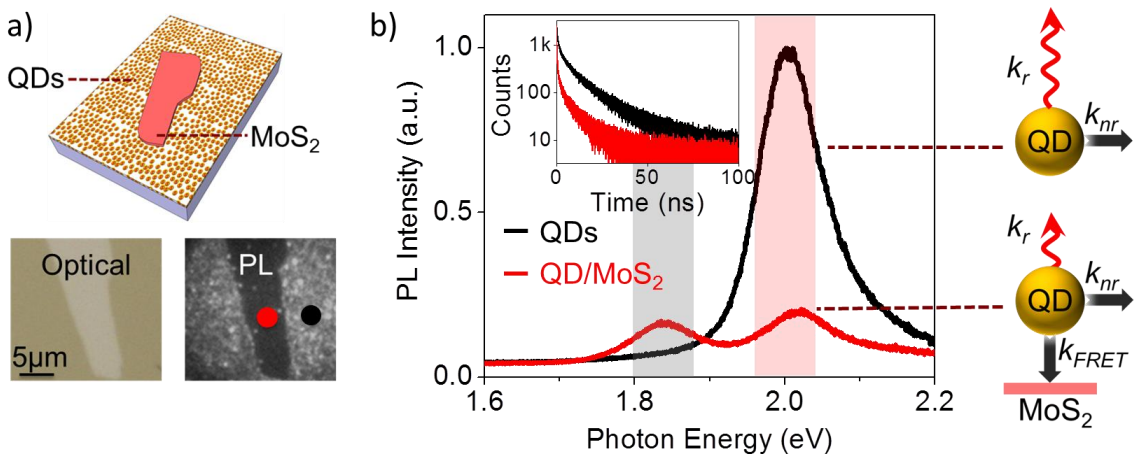

**Figure 3 (a)** Ungated $MoS_2$/QD device along with its optical (left) and photoluminescence (right) images. PL image was recorded using a band-pass filter (605nm-615nm) only transmitting QD emission. **(b)** PL spectra and time-resolved PL (Inset) of QD/$MoS_2$ hybrid (red) and of bare QD film (black). The spectra were recorded from the same device shown in Fig. 3a at positions marked by red and black circles. The schematic on the right illustrates FRET between a QD and $MoS_2$.

not differ significantly from that of same QDs in solution (Fig 2c, black dotted line). This suggests that the QDs are not chemically modified during the process of self-assembly and that the interactions between QDs are negligible. Each QD in the film can therefore be treated as a single emitter.

Finally, we mechanically transferred a monolayer MoS$_2$ onto QDs using fabrication techniques developed for 2D heterostructures.[25, 26] Several experimental tests described below confirm that such transfer does not perturb the QD layer.

**Results and Discussion**
**Experimental evidence of FRET**

A typical sample along with its optical and PL image is shown in Fig. 3a. This sample can be considered ungated ($V_g=0$) compared to electrostatically gated devices studied further. Both the PL image and PL spectra (Fig. 3a,b) indicate strong suppression of photoluminescence for the QDs that are close to MoS$_2$. To quantify this effect, we introduce the quenching factor $Q = I_{QD}/I_{QD/MoS_2}$. Here $I_{QD/MoS_2}$ is the height of the QD photoluminescence peak at 2.02eV for the hybrid QD/MoS$_2$ device (acquired at a point marked red in Fig. 3a), and $I_{QD}$ is the height of the same peak from QDs away from MoS$_2$ (acquired at a point marked black in Fig. 3a). We calculate $Q(0V)\sim 4.8$ from the data shown in Fig. 3b. We also observed that the lifetimes of QDs reduce by a similar amount due to the presence of MoS$_2$, $\tau_{QD}/\tau_{QD/MoS_2}\sim 4.4$ (Fig. 3b, Inset). At the same time, the position of the PL peak due to QDs remained virtually unchanged at about ~2.02eV (Fig. 3b). This indicates that the QDs are not chemically or mechanically perturbed by MoS$_2$.

The quenched PL and decreased lifetimes indicate the opening of an additional non-radiative relaxation channel for the QDs next to MoS$_2$. We attribute this pathway to FRET. Strong spectral overlap between the emission spectrum of QDs and B-peak in absorption of MoS$_2$ coupled with very small QD/MoS$_2$ separation should, according to Eq. (1), lead to large $k_{FRET}$. Prior experiments on similar QDs next to 2D systems (graphene, MoS$_2$) arrived at a similar conclusion.[27, 28]

We confirmed that mechanisms other than FRET are not responsible for observed changes in PL in our devices. In principle, charge transfer between QDs and MoS$_2$ can also lead to non-radiative relaxation.[29-31] For our experiments we intentionally chose core-shell QDs with strong electron-hole pair confinement and long ligands.[24] Charge transfer in such core-shell QDs is likely inefficient or absent.[32] To further exclude the contribution of charge transfer, we fabricated devices with a spacer layer (5-15nm of SiO$_2$) inserted between QDs and MoS$_2$. Despite large MoS$_2$/QD separation, we observed significant quenching in PL of QDs atop of MoS$_2$ (Supporting Information, S2). Such quenching can only be attributed to long-range FRET, as short-range charge transfer should be fully suppressed in spacer devices.[33] In addition, charge transfer is conclusively ruled by the optoelectronic measurements described in the last section of the manuscript. It is also feasible that dielectric screening due to MoS$_2$ could affect the intensity of QD photoluminescence. To exclude this possibility, we fabricated devices where hBN, an optically transparent insulator, is transferred onto QDs instead of $MoS_2$ (Supporting Information, S3). While hBN has a dielectric constant ε ~ 4–7, [34] similar to that of monolayer $MoS_2$,[35] we did not observe any spectral changes or quenching for QDs in hBN/QD devices. This confirms that the QDs are not affected by dielectric screening due to neighboring materials. This also rules out the possibility of mechanical or chemical changes to the QD layer during the transfer procedure.

The $QD/MoS_2$ FRET rate was estimated from measured suppression of QD photoluminescence and lifetimes. The intensity of QD photoluminescence depends on radiative ($k_r$) and non-radiative ($k_{nr}, k_{FRET}$) decay rates:

$$I_{QD} \sim \frac{k_r}{k_r + k_{nr}} = k_r \tau_{QD},$$

$$I_{QD/MoS_2} \sim \frac{k_r}{k_r + k_{nr} + k_{FRET}} = k_r \tau_{QD/MoS_2} \quad (2)$$

In these equations, the lifetime of a QD is expressed as an inverse of the sum of radiative and non-radiative rates, and $k_r$ is assumed to be unaffected by the environment. Equation (2) confirms that near-equal suppression of QD lifetime and PL intensity observed in our experiments is an expected consequence of FRET. From the measured PL quenching $Q \sim 4.8$, using equation (2) we determined $k_{FRET} = (Q-1)/\tau_{QD} \sim (1.1 \pm 0.2) \times 10^9 s^{-1}$. Importantly, this rate corresponds to lifetime ~1ns, shorter than the intrinsic QD lifetime of ~3ns. From measured $Q$ and assuming separation distance between QD-core and $MoS_2$ ~3.5nm (Fig 2b, inset), we evaluate FRET radius $R_0$~5nm.

**Electrical modulation of FRET**

Finally, we examined gate-induced modification of the optical properties of $QD/MoS_2$ devices. To enable such a study, we used fabrication described previously, but with $MoS_2$ transferred on top of pre-patterned gold electrodes. An optically transparent solid electrolyte was then deposited onto $MoS_2$ (Fig. 4a, see "Methods" for details). This configuration allows us to vary the carrier density inside $MoS_2$ while being able to perform optical measurements. It is also important to note that electric field is near-absent at the location of QDs and cannot affect their photoluminescence directly.[36] Although very high carrier densities, $n \sim 10^{14}$ $cm^{-2}$, can be reached with electrolyte gates (Supporting Information, S4),[37] our devices require much smaller densities, $n \sim 10^{13}$ $cm^{-2}$, and efficiently operate at low gate voltages (-2V<$V_g$<2V). Overall, we fabricated and measured 4 devices including the representative device shown in Fig. 4a.

With increased electron doping (positive $V_g$), we observed a well-known suppression of the PL peak[2,3] due to $MoS_2$ at 1.88eV as discussed earlier (Fig. 1b and Supporting Information Fig. S4b). On the other hand, photoluminescence of QDs at ~2.02eV strongly increases with $V_g$ (Fig. 4b). In our best device, we observed up to ~75% modulation of the QD photoluminescence intensity for $V_g$ between -2V and 2V. This effect is reproducible for all measured devices and is stable over multiple sweeps of $V_g$ (Fig. 4b, Inset).

We attribute the modulation of PL to gate-induced modulation of the FRET rate $k_{FRET}$. Indeed, as discussed above, optical absorption

$\alpha(\lambda)$ of MoS$_2$ is strongly changing with $V_g$ at 2.05 eV, the energy corresponding to QD emission (Fig. 1c). According to the equation (1), changes in $\alpha(\lambda)$ should lead to modulation of the FRET rate, and hence QD PL intensity.

Our next goal is to understand the relationship between FRET modulation and MoS$_2$ absorption. In a separate measurement on a device without a QD layer, we used

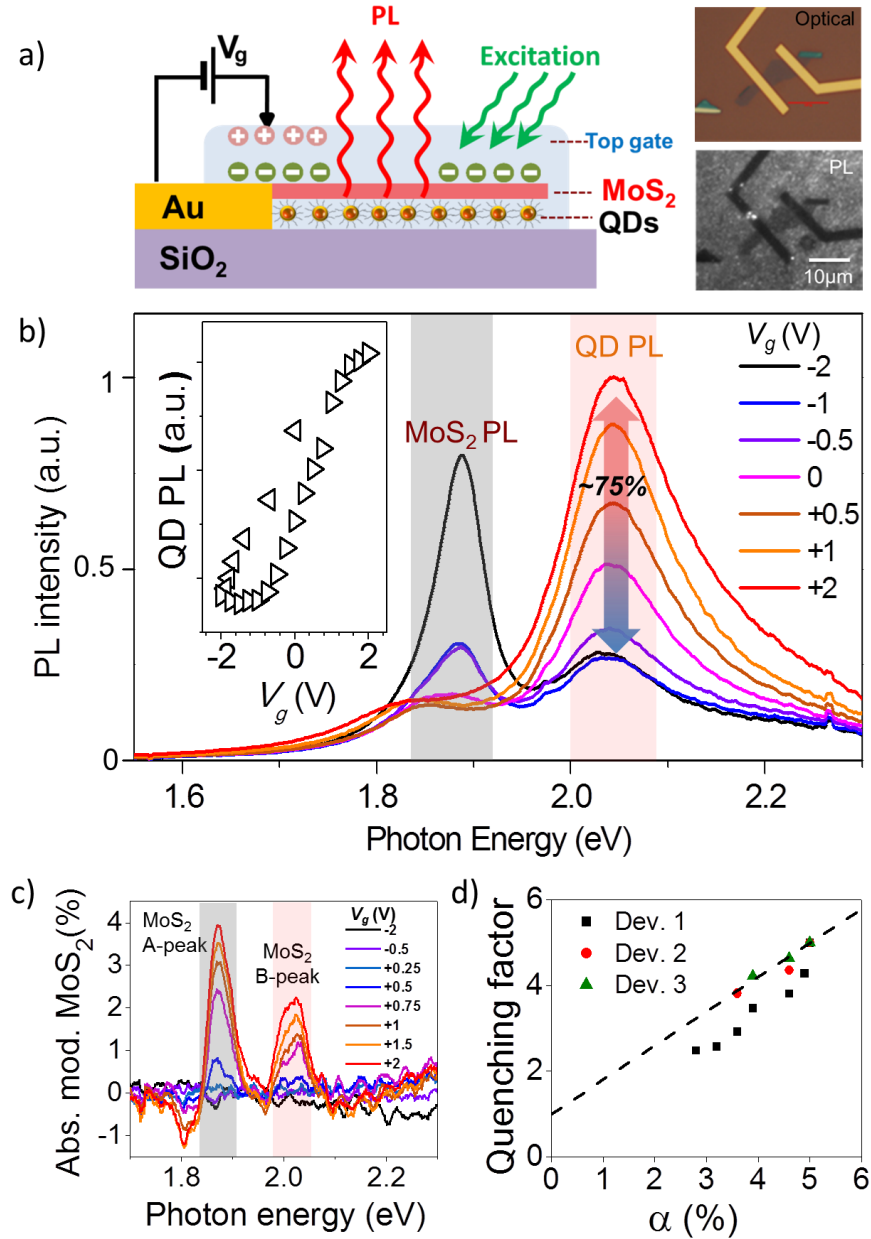

**Figure 4 (a)** Device schematic of electrolyte gated QD/MoS$_2$ hybrid. Optical and photoluminescence images of an electrically contacted QD/MoS$_2$ device. **(b)** PL spectra of a QD/MoS$_2$ device at different $V_g$. Inset: QD photoluminescence intensity *vs.* $V_g$ during a back-and-forth sweep between +2V and -2V. **(c)** Transmittance modulation of MoS$_2$ The dip at ~1.8eV is likely related to charged exciton absorption. **(d)** From the measured $Q$ vs. $V_g$ from (a) and $\alpha$ vs. $V_g$ from (b), a single parametric $Q(\alpha)$ plot was created. Since the transmission of MoS$_2$ is only reliably determined for $V_g > 0$, only these points were used in the plot (details in Supporting Information, S6).

confocal transmission microscopy to record gate-induced transmittance modulation of MoS$_2$ defined as

$$M = \left(I(\hbar\omega,\ V_g) - I(\hbar\omega,\ 0V)\right)/I(\hbar\omega,\ 0V).$$

Here $I(\hbar\omega, V_g)$ is the intensity of light transmitted through MoS$_2$ at photon energy $\hbar\omega$ and gate voltage $V_g$. We use transmittance modulation as a proxy measurement for far-field absorption which is otherwise hard to assess via conventional differential reflectivity measurements for our device geometry. A simple estimate yields $\alpha(V_g) = \alpha(V_g = 0V) - M(V_g)$ (Details in Methods and Supporting Information, S5). Within our gating range we observe only ~2% modulation of MoS$_2$ transmittance at ~2.05eV (Fig. 4c), much smaller than ~75% modulation in QD photoluminescence.

We devised a simple model relating near-field FRET rate and quenching factor to far-field absorption of MoS$_2$. The normalized emission spectrum of an individual QD centered at wavelength $\lambda$ is narrow compared to the relatively broad absorption features of MoS$_2$.[38] In this situation, equation (1) can be simplified to

$$k_{FRET} \sim \frac{1}{d^4}\alpha(\lambda, V_g).$$

Combining this with equation (2), we obtain the following expression for the quenching factor $Q$:

$$Q(\lambda, V_g) = \frac{\tau_{QD}}{\tau_{QD/MoS_2}} = 1 + \tau_{QD}k_{FRET}$$
$$= 1 + A\alpha(\lambda, V_g). \quad (3)$$

Here $A \sim \frac{\tau_{QD}}{d^4}$ is a proportionality constant relating the quenching factor to absorption of MoS$_2$. From experimentally measured $Q(V_g=0)\sim 5$ (Fig. 3b) and $\alpha(V_g=0)\sim 5\%$ (Fig. 1b) at $\lambda$=610nm (QD emission peak), we find $A = (Q(0) - 1)/\alpha(0)\sim 80$. The large value of $A$ translates to large electrical modulation of PL of the QDs. To check the validity of our model, we plotted experimentally acquired values of $Q$ and $\alpha$. The measured $Q(\alpha)$ along with the prediction of equation (3) (dashed line) are plotted in Fig. 4d. The agreement between the experimental data and our model confirms that the observed modulation of QD photoluminescence is a consequence of electrical modulation of FRET. From Fig. 4b (inset) and equation (3) we also find that the FRET rate changes from $2.8 \times 10^9 s^{-1}$ to $0.5 \times 10^9 s^{-1}$ within our gating range.

We devised additional control experiments to further confirm that the observed PL modulation is related to gate-induced changes in excitonic absorption of two-dimensional semiconductors and not to other mechanisms. We fabricated one device where MoS$_2$ is substituted by a monolayer of graphene and another QD/MoS$_2$ device with different CdSSe QDs emitting at ~2.2eV, not in resonance with MoS$_2$ absorption peaks. In contrast to the devices discussed above (e.g. in Fig. 4), in both of these samples optical absorption of the 2D material is gate-independent at the QD emission wavelength (Fig. 4c and Supporting Information, S7). As expected, since FRET modulation is spectrally selective, we did not observe any gate-dependent changes of the QD photoluminescence in either device in the range of gate voltage between -3V and 3V. Finally, we fabricated a device with QDs emitting at ~2.4eV, but with a different 2DSC, WS$_2$, instead of MoS$_2$. Large and clear modulation of QD PL is observed in this device since the gate-dependent excitonic peaks of WS$_2$ (A-peak: 2.0eV, B-peak: 2.4eV)[39] are in resonance with the QD emission peak (Supporting Information, S8).

These observations confirm that PL of QDs is only affected by the absorbance of a 2D material at relevant frequencies and not just its carrier density. We therefore strengthen our claim that charge transfer between MoS$_2$ and QDs is either absent or does not depend on gate voltage. The lack of PL modulation in QD/graphene devices further highlights the advantage of 2DSCs for modulation of QDs in the visible (as opposed to IR[12, 13]) range. Furthermore, we see that QD/2DSC hybrids can be used for selective modulation of QDs emitting at different wavelengths.

In summary, we demonstrated electrical control of the near-field energy transfer between QDs and two-dimensional semiconductors (MoS$_2$, WS$_2$). We found that it is related to modulation of excitonic absorption of 2D semiconductors, and achieved ~75% modulation of QD photoluminescence in the visible range. It is instructive to compare our approach to other existing schemes to control photoluminescence of QDs via electrical signals. Some of the existing schemes utilize electrochemical injection of charge carriers into QDs,[30, 31] electron-hole dissociation under applied electric fields,[40] or controlled Stark shifts.[41] In all of these schemes, electrical fields are applied directly to the QDs. In our approach the electric field changes the parameters of a two-dimensional semiconductor and is absent at the location of QDs.[36] We do not expect electrochemical modification of QDs. The operating principle of our scheme – electrical control over the QD/2DSC FRET rate – can be extended to other nanoemitters. Finally, QDs emitting at different wavelengths over the visible and IR ranges can be modulated by choosing two-dimensional semiconductors with varied bandgaps (e.g.: WSe$_2$, WS$_2$, MoSe$_2$).

We envision several potential improvements in our system. FRET efficiency, and hence the efficiency of PL modulation, can be increased by reducing the distance between QDs and 2DSCs (equations (1) and (3)). This can be achieved by either reducing QD shell-size or by shortening QD ligands. Additionally, 2DSCs could be gated more efficiently using ultrathin gate dielectrics. The advances in CVD growth[42, 43] of 2DSCs could lead to inexpensive fabrication of large-scale QD/2DSC hybrids. Overall, QD/2DSCs hybrids could be used as efficient and electrically tunable light sources operating anywhere in the visible to IR spectral range. Potential applications for such devices range from solid-state lighting and high-resolution passive ("e-ink") displays to biosensors.

**Methods**

**Synthesis of CdS$_x$Se$_{1-x}$ Graded Alloy Quantum Dots.** This one-pot synthetic procedure is based on a method published recently by Harrison *et al.*[19] First, 1 mmol CdO (0.128 g), 1.3 mL oleic acid (HOA), and 20 mL 1-octadecene (ODE) were heated to 100°C under vacuum for 10 minutes, and subsequently purged with Ar. The temperature was increased to 260°C and the conversion of red CdO to colorless Cd-oleate was monitored to completion, after which the reaction temperature was reduced to 220°C. Solutions of S:Tributyl phosphate (0.75 M) and Se:Tributyl phosphate (0.75 M) in ODE were prepared separately and 0.8 mL aliquots of each were pulled into the same syringe. The S/Se aliquot was swiftly injected into the Cd-oleate flask at 220°C and the reaction was allowed to proceed for 2hrs. The nanocrystals were cooled and precipitated with a 3:1 mixture of butanol and ethanol, resuspended in toluene, and precipitated twice more with pure ethanol. After being finally suspended in toluene, the nanocrystals were passed through a 0.45μm filter and stored.

**QD/MoS$_2$ device fabrication.** Cr/Au (2nm/30nm) electrodes were deposited on SiO$_2$ substrates. The substrates were then cleaned in a piranha solution (1:3 H$_2$O$_2$:H$_2$SO$_4$) for 1 hour, made hydrophilic through O$_2$ plasma treatment (30s), and functionalized in 1mM solution of (3-Mercaptopropyl) trimethoxysilane in hexane for 10 min. Functionalized substrates were washed in a hexane bath for 1 min, rinsed in isopropanol, and blow-dried. To assemble a uniform film of QDs, functionalized substrates were placed into 5mg/ml solution of CdSSe for 30mins and rinsed gently afterwards with toluene. To transfer MoS$_2$ onto QDs, we followed the recipe developed by Zomer et al.[26] We spun Elvacite polymer (~1μm thick) onto PDMS/clear Scotch tape sandwich structure. The structure was baked at 90°C for 5mins. Monolayer MoS$_2$ was exfoliated onto Elvacite and verified using optical microscopy and Raman spectroscopy. MoS$_2$ was aligned with Au electrodes, brought into contact with QD films and baked at 120°C. The PDMS/polymer layer was then mechanically separated from the MoS$_2$/QD stack. To remove the polymer residues, the MoS$_2$/QD stack was soaked in acetone for 15 min. Finally, we created the solid electrolyte gate by placing a drop of CsClO$_4$ salt in poly(ethylene) oxide (PEO) matrix dissolved in acetonitrile and drying it for 2hrs at room temperature. A second gate electrode close to MoS$_2$ was used to contact the solid electrolyte.

**PL measurements.** PL spectra were recorded at ambient conditions using a Thermo Scientific DXR Raman microscope with a 100μW, 532nm (~2.3eV) laser as an excitation source. MoS$_2$ was electrically gated using a Keithley 2400 sourcemeter connected to the solid electrolyte. PL modulation of MoS$_2$ was used to confirm gating efficiency. PL images were collected using a conventional fluorescence microscopy setup with a 605-615nm bandpass filter and green light (530–590 nm) excitation.

**Time resolved PL measurements.** PL lifetimes of QDs were recorded using a modified version of a home-built confocal microscope described previously.[44] A 400 nm pulsed beam with a repetition rate of 250kHz was reflected from a 410nm long-pass dichroic filter (Omega Optics 3RD410LP) and focused through a water immersion objective to a confocal spot on the QD layer of the fabricated devices. PL was collected through the objective and subsequently passed through the dichroic filter and a 610 ± 5nm bandpass filter to select for QD PL. The QD photoluminescence was then focused onto the array of a single photon avalanche diode (Micro Photon Devices PDM series SPAD). Lifetime data was collected in the form of single photon events *via* a time correlated single photon counting (TCSPC) correlator (PicoHarp 300) with a time resolution of 4ps. Time-resolved PL from $MoS_2$ was measured using a grating spectrometer (Acton) coupled to a streak camera system (Hamamatsu). The second harmonic of a femtosecond Ti:sapphire laser with 450 nm pump pulses, 100fs in duration was used for excitation. Two-dimensional spectrograms were acquired in photon-counting mode with 2nm spectral resolution and a minimum 3ps temporal resolution. Time-resolved PL spectra were fitted by a tri-exponential function and lifetimes were estimated as weighted averages of three decay rates.

**Absorption/transmittance modulation measurements.** Standard differential reflectivity measurements could not be performed on our samples due to the non-uniformity of the solid electrolyte layer. Instead, we used confocal transmission microscopy to determine absorbance/transmittance of gated $MoS_2$ devices on transparent glass substrates. A broad (~1mm) light beam from a fiber-coupled halogen light source was used to illuminate our sample. Light passed through the sample was collected through a 40X objective and was further magnified ~10 times and focused on a screen with a ~0.5mm diameter pinhole. The pinhole blocks the light from the rest of the sample while transmitting light that passes through $MoS_2$. The spectrum of the transmitted light as a function of gate voltage was recorded using Shamrock 303i spectrometer. We note that due to the low quantum yield of $MoS_2$,[6] its PL cannot interfere with our absorption measurements. Differential transmittance measurements of $MoS_2$ devices on glass without the solid electrolyte layer (Fig. 1b) were obtained using the same technique.

**Supporting Information**
Qualitative discussion of 0D/2D FRET. PL measurement of $MoS_2$/Spacer/QD devices. PL Measurement of QD/hBN devices. Measurements of carrier density *vs*. $V_g$ for a graphene Hall-bar device gated using solid electrolyte. PL spectra *vs*. $V_g$ for a $MoS_2$ device gated using solid electrolyte. Detailed discussion and schematic of optical transmittance measurements. Detailed description of $Q(\alpha)$ measurements. PL spectra *vs*. $V_g$ for control QD/graphene and QD/$MoS_2$ devices. PL spectra *vs*. $V_g$ for green QD/$WS_2$, green QD/$MoS_2$ and red QD/$MoS_2$

**Corresponding Author**

*Email: kirill.bolotin@vanderbilt.edu

**Acknowledgements**
We acknowledge useful discussions with Kirill Velizhanin. The work was primarily supported by the Office of Naval Research award N000141310299 with additional funding from NSF EPS-1004083, NSF CBET-1134509 and NSF DMR-1056859.

# Supporting Information

## S1. Supplementary discussion: FRET between 0D and a 2D system

In general, the rate of Forster Resonant Energy Transfer ($k_{FRET}$) between two quantum dots follows from the Fermi's golden rule and is well-known (Reference 1 of the main manuscript):

$$k_{FRET(0D-0D)} \sim |\vec{E}|^2 \int_0^\infty \varepsilon_A(\lambda) f(\lambda) \lambda^4 d\lambda.$$

Here $|\vec{E}|^2$ is the square of electric field created by a QD dipole at the position of the other QD, $f(\lambda)$ is the normalized emission spectrum of the donor QD as a function of wavelength $\lambda$, and $\varepsilon_A(\lambda)$ is the acceptor molar extinction coefficient. Since dipole field decays with distance as $r^{-3}$, $|\vec{E}|^2$ is proportional to $r^{-6}$. Therefore, $k_{FRET(0D-0D)}$ has the same distance dependence.

For the case of FRET between a 0D and a 2D system, we can formally split a 2D material into a 2D array of point-like absorbers and then integrate $k_{FRET(0D-0D)}$ over 2D material area. If the 2D material is located in oXY plane and a QD is elevated above that plane by distance $d$, the distance $r$ between QD and the point on the plane can be written as $r = \sqrt{d^2 + x^2 + y^2}$. Then integration over area yields:

$$k_{FRET(0D-2D)} \sim \int_{Area} d(Area)\, k_{FRET(0D-0D)} \sim \int_{Area} \frac{dxdy}{\sqrt{d^2 + x^2 + y^2}^{\,6}} \int_0^\infty \varepsilon_A(\lambda) f(\lambda) \lambda^4 d\lambda.$$

Replacing absorptivity $\varepsilon_A(\lambda)$ by 2D absorption coefficient $\alpha(\lambda)$ and performing simple integration over area we obtain the expression used in the main text:

$$k_{FRET} \sim \frac{1}{d^4} \int_0^\infty \alpha(\lambda) f(\lambda) \lambda^4 d\lambda.$$

## S2. MoS$_2$ – Spacer – QD Device

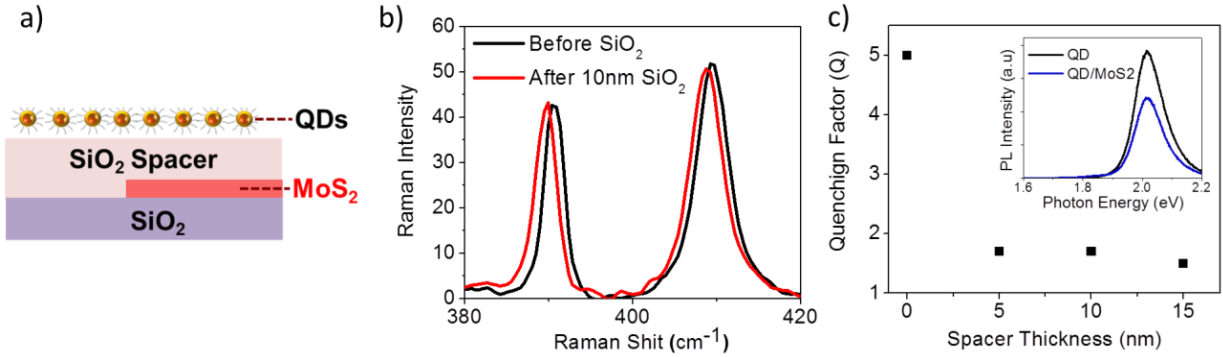

**Supplementary Figure S2 (a)** Device schematic for MoS$_2$/spacer/QD device. **(b)** Raman spectra of MoS$_2$ before (black) and after (red) SiO$_2$ spacer deposition. **(c)** QD quenching factor vs. spacer thickness. Inset: PL spectra of QD and QD/MoS$_2$ for 15nm spacer device.

## S3. hBN – QD Device

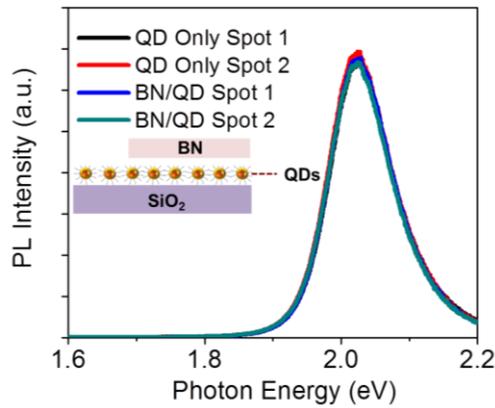

**Supplementary Figure S3** Photoluminescence spectra for QDs covered by hBN (blue and green curves) and QDs away from hBN (black and red curves). Inset: schematic of a hBN/QD device.

## S4. Solid electrolyte characterization

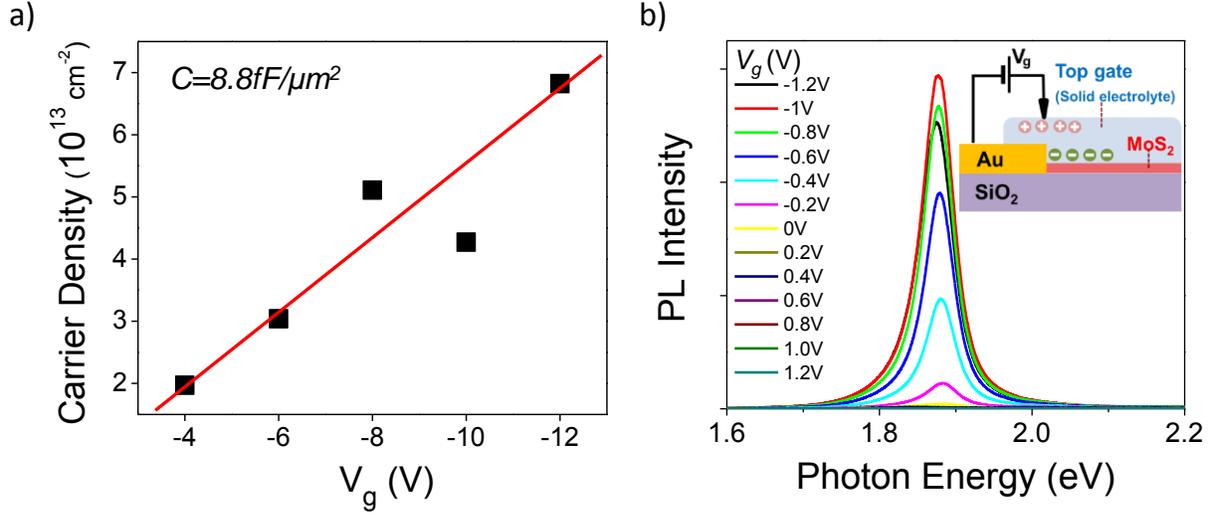

**Supplementary Figure S4 (a)** The efficiency of the solid electrolyte gating approach was estimated using a separate graphene Hall-bar device covered by the same electrolyte used in QD/MoS$_2$ devices. Carrier density *n vs.* gate voltage $V_g$ in that device was determined via Hall measurements. From a fit (red line) to the acquired $n(V_g)$ data (black symbols), we extract an estimate for the capacitance of the polymer electrolyte, C~8.8fF/$\mu$m$^2$. **(b)** PL modulation of a representative gated MoS$_2$ device without the QD layer. Schematic of the device is shown in the inset.

## S5. Confocal transmission microscopy

The absorption spectrum of MoS$_2$ could not be obtained directly from standard differential reflectivity measurements for our electrolyte gated MoS$_2$ samples. This is due to the non-uniformity of the solid electrolyte layer. Instead, we used confocal transmission microscopy to determine transmittance modulation of gated MoS$_2$ devices on transparent glass substrates (Fig. S5). Transmittance modulation is defined as $M = \left(I(\hbar\omega,\ V_g) - I(\hbar\omega,\ 0V)\right)/I(\hbar\omega,\ 0V)$, where $I(\hbar\omega, V_g)$ is the intensity of light transmitted through MoS$_2$ at photon energy $\hbar\omega$ and gate voltage $V_g$. Transmittance modulation is closely related to absorption modulation. We can rewrite the definition of *M* as

$$M = \frac{\frac{I(\hbar\omega,\ V_g)}{I_0(\hbar\omega)} - \frac{I(\hbar\omega,\ 0V)}{I_0(\hbar\omega)}}{\frac{I(\hbar\omega,\ 0V)}{I_0(\hbar\omega)}}.$$

Here $I_0(\hbar\omega)$ is the intensity of the incident light. Since $I(\hbar\omega,\ V_g)/I_0(\hbar\omega) = 1 - \alpha(\hbar\omega,\ V_g)$, we get:

$$M = \frac{\alpha(\hbar\omega,\ 0V) - \alpha(\hbar\omega,\ V_g)}{1 - \alpha(\hbar\omega, 0V)}.$$

Since MoS$_2$ absorption is small (~5%) in our wavelength region, $M \approx \alpha(\hbar\omega, 0V) - \alpha(\hbar\omega, V_g)$ or $\alpha(\hbar\omega, V_g) = \alpha(\hbar\omega, V_g = 0V) - M(\hbar\omega, V_g)$. Therefore, we can estimate $\alpha(\hbar\omega, V_g)$ from measured $M$ and $\alpha(\hbar\omega, V_g = 0V)$~5% obtained from an unbiased MoS$_2$ flake before deposition of solid electrolyte.

Experimentally, a broad (~1mm) light beam from a fiber-coupled halogen light source was used to illuminate our sample. Light passed through the sample was collected through a 40X objective and was further magnified ~10 times and focused on a screen with a ~0.5mm diameter pinhole. A magnified image of the device was projected on the screen. The pinhole was adjusted to block the light from the rest of the sample while transmitting light that passes through MoS$_2$. The spectrum of the transmitted light as a function of gate voltage was recorded using Shamrock 303i spectrometer.

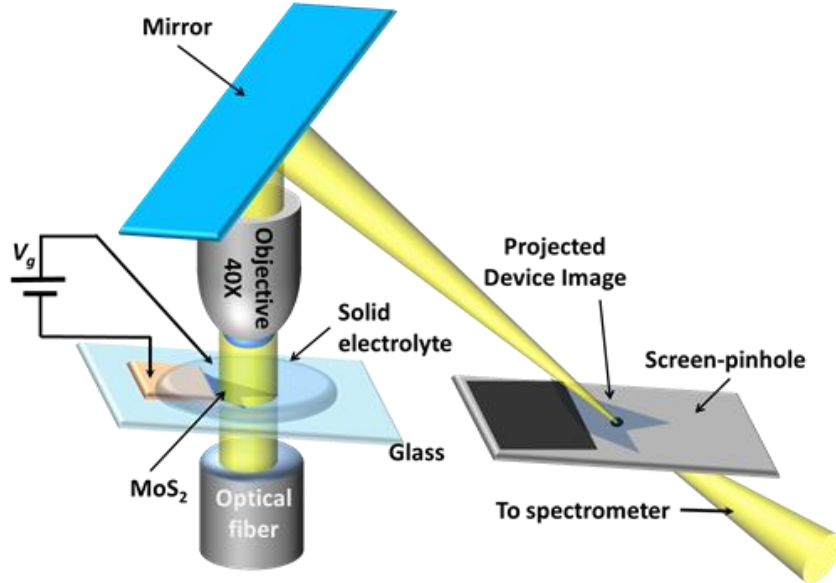

**Supplementary Figure S5** Schematic of measurement set up for transmission microscopy.

**S6. Supplementary Discussion: Detailed description of $Q(\alpha)$ measurement**

The data in the $Q(\alpha)$ parametric plot (Fig. 4d, main text) were obtained as follows. A QD/MoS$_2$ device was used for measurements of the quenching factor $Q$ vs. $V_g$. However, absorption of the MoS$_2$ layer, $\alpha_{MoS_2}$, could not be determined in the same device due to the strong background absorption of the QDs. For that reason, a separate MoS$_2$-only device without QDs was used for $\alpha_{MoS_2}$ vs. $V_g$ measurements (Fig. 4c of the main text, Supporting information S5). However, interpretation of $Q(\alpha)$ data is complicated by the difference of the intrinsic doping levels of MoS$_2$ between QD/MoS$_2$ and MoS$_2$-only devices. Indeed, at $V_g=0$ we observed reduced PL due to MoS$_2$ (peak at ~1.9eV) in QD/MoS$_2$ as compared to MoS$_2$-only devices (Fig. S4b). Since PL of MoS$_2$ can be used as a proxy for free carrier density (Fig. 1b, Fig. S4b, Fig.S6a), this observation suggests that the intrinsic doping level of MoS$_2$ in MoS$_2$-only devices is *lower* than that of MoS$_2$ in QD/MoS$_2$ devices. Moreover, the observation of near-absent absorption modulation for MoS$_2$ in MoS$_2$-only devices for $V_g<0$ (as compared to strong absorption modulation for $V_g>0$) suggests that the density of free carriers in that device approaches ~0 at $V_g=0$ (Fig. 4c). In contrast, robust changes of MoS$_2$ and QD photoluminescence in QD/MoS$_2$ devices (Fig. 4b, main text) hint that the density of free carriers is changing throughout our gating range and the Fermi level always stays within the conduction band. In Fig.S6b, we illustrated the proposed Fermi level positioning between MoS$_2$ and QD/MoS$_2$ devices due to difference in intrinsic doping levels.

Because of the difference in the intrinsic doping levels, we have to be careful in relating the experimentally measured $\alpha_{MoS_2}$ to the analysis of QD/MoS$_2$ devices. For $V_g>0$, the Fermi level of MoS$_2$ in both MoS$_2$-only and QD/MoS$_2$ devices is in the conduction band and the absorption of MoS$_2$ in both devices changes similarly. On the other hand, when $V_g<0$, the Fermi level of MoS$_2$ in MoS$_2$-only devices is shifted below the conduction band edge. In that case, the density of free carriers and hence $\alpha_{MoS_2}$ are nearly $V_g$-independent (Fig. S6b). At the same time, the absorption of MoS$_2$ in QD/MoS$_2$ devices strongly changes with $V_g$. This means $\alpha_{MoS_2}$ in QD/MoS$_2$ and MoS$_2$-only devices are only close when $V_g>0$. Because of that, the data in Fig. 4c of the main text are only plotted in that range.

For completeness, we show $Q(\alpha)$ for the entire range from -2V to 2V in the Figure S6c. The discussion above shows that the deviation from linear dependence for $V_g<0$ is caused by inaccuracy in measured MoS$_2$ absorption in that voltage range.

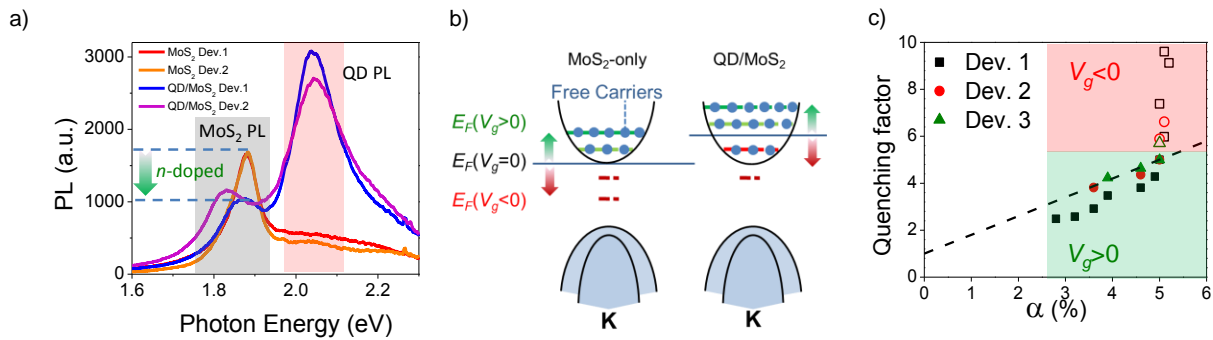

**Supplementary Figure S6 (a)** Photoluminescence spectra of two different MoS$_2$-only devices without QDs, and of two different QD/MoS$_2$ devices. **(b)** Proposed Fermi level ($E_F$) positioning of MoS$_2$ in MoS$_2$-only and QD/MoS$_2$ devices. **(c)** $Q(\alpha)$ plot including the data for $V_g<0$.

## S7. Off-peak QD/MoS$_2$ and QD/Graphene Devices

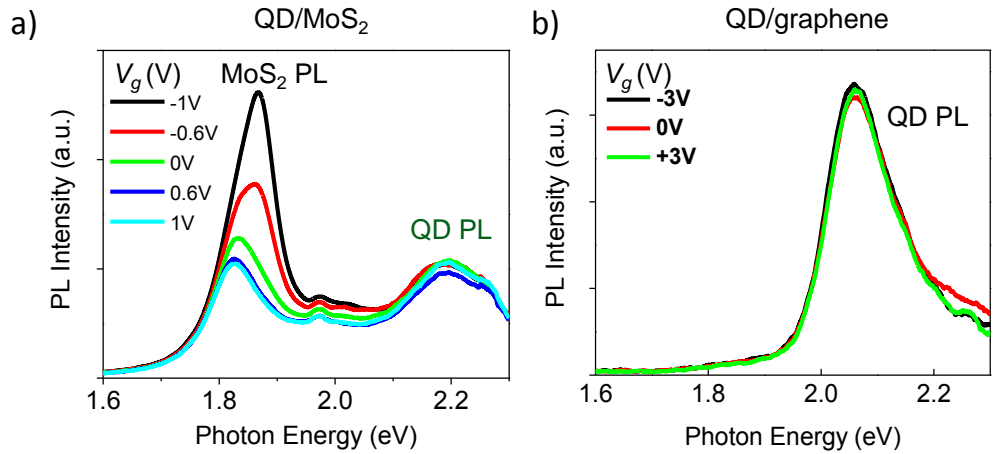

**Supplementary Figure S7** To check possible contribution of charge transfer to QD PL modulation in QD/MoS$_2$, two additional types of devices were fabricated. In the first type of device **(a)** we used CdSSe QD with the emission peak at ~2.2eV (away from the excitonic absorption peak of MoS$_2$) to make hybrid QD/MoS$_2$ devices. In the second type of device **(b)**, same QDs as in the rest of the manuscript (emission peak at 2.02eV) were used, but MoS$_2$ was substituted by monolayer graphene. In both devices, optical absorption of the 2D material was constant at relevant QD emission energies. PL spectra were recorded while varying the gate voltage $V_g$ for both (a) QD/MoS$_2$ and (b) QD/graphene devices. In both cases, we observed no changes in the PL at the emission wavelength of the QDs (2.2 eV in (a) and 2.02eV in (b)). This indicates that electrical modulation of the PL for the QDs used in the manuscript is due to changes in excitonic absorption of MoS$_2$ and not just due to changes in its carrier density.

## S8. Spectrally selective tuning of QDs using WS$_2$ and MoS$_2$

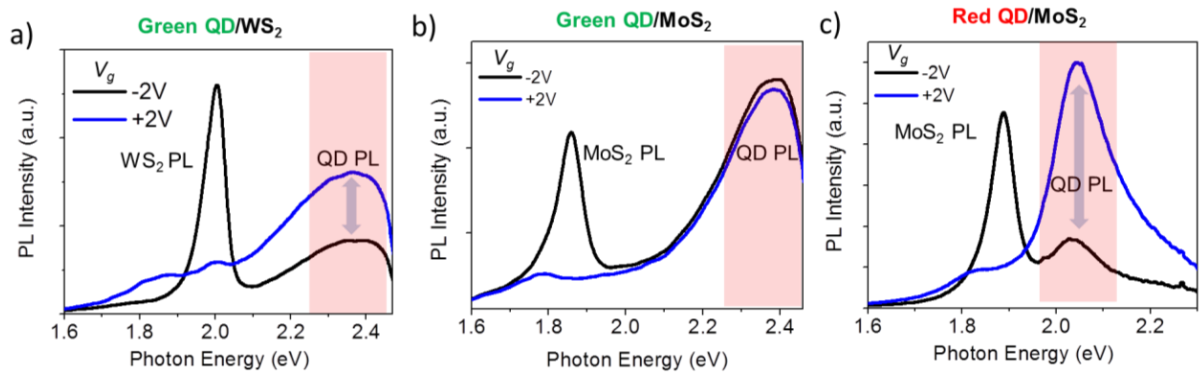

**Supplementary Figure S8.** **(a)** QD/WS$_2$ devices with QDs emitting at ~2.4eV (green color). The emission peak of these QDs **is** in resonance with excitonic absorption peak of WS$_2$ **(b)** QD/MoS$_2$ devices with QDs emitting at ~2.4eV (green color). The emission peak of these QDs **is not** in resonance with MoS$_2$ excitonic absorption peak **(c)** QD/MoS$_2$ devices with QDs emitting at ~2.02eV (red color). The emission peak of these QDs **is** in resonance with MoS$_2$ excitonic absorption.